# Investigating the time-dependent behaviour of Boom clay under thermo-mechanical loading

Yu-Jun CUI[1], Trung Tinh LE[1], Anh Minh TANG[1], Pierre DELAGE[1], Xiang Ling LI[2]
*1 : Ecole des Ponts - ParisTech, UR Navier/CERMES, 6 & 8 av Blaise Pascal, Champs-sur-Marne, F 77455 Marne-la-Vallée Cedex 2*
*2: EURIDICE Group, c/o SCK/CEN, Boeretang 200, 2400 Mol, Belgium*

**Corresponding author:**

Prof. Yu-Jun CUI
ENPC/CERMES
6 et 8, av. Blaise Pascal
Cité Descartes, Champs-sur-Marne
77455 Marne la Vallée cedex 2
France
Tel: 33 1 64 15 35 50
Fax: 33 1 64 15 35 62
E-mail: cui@cermes.enpc.fr

**Abstract**
Boom clay, a stiff clay, has been selected as a potential host formation for the geological disposal of radioactive waste in Belgium. The underground research facility HADES has been constructed to enable various in-situ experiments to be performed on Boom clay so as to study the feasibility of High Level radioactive Waste (HLW) disposal and to provide reliable data on the performance of Boom clay as a host formation. Among the various laboratory studies performed on samples extracted from the HADES facility to investigate the Thermo-Hydro-Mechanical (THM) behaviour of Boom clay, relatively few were devoted to the time dependent behaviour, limiting any relevant analysis of the long-term behaviour of the disposal facility. The present work aims at investigating the time-dependent behaviour of Boom clay under both thermal and mechanical loading. High-pressure triaxial tests at controlled temperatures were carried out for this purpose. The tests started with constant-rate thermal and/or mechanical consolidation and ended with isobar heating and/or isothermal compression at a constant stress rate or by step loading. The results obtained confirmed the effect of the overconsolidation ratio (OCR) on the thermal volume changes (i.e. thermal dilation under high OCRs and thermal contraction with OCR close to unity). Significant effects of temperature as well as of compression and heating rates were also observed on the volume change behaviour. After being loaded to a stress lower than the pre-consolidation pressure (5 MPa) at a low temperature of 25°C and at a rate lower than 0.2 kPa/min, the sample volume changes seemed to be quite small, suggesting a full dissipation of pore water pressure. By contrast, after being subjected to high loading and heating rates (including step loading or step heating), the volume changes appeared to be significant, particularly in the case of stresses much higher than the pre-consolidation pressure. Due to low permeability, full consolidation of Boom clay required a long period of time and it was difficult to distinguish consolidation and creep from the total volume change with time.

# 1. Introduction

Boom clay, a tertiary clay formation located at a depth of between 190 m and 290 m in Mol (Belgium) was selected as a potential host formation for the disposal of High Level and Long lived radioactive Waste (HLLW). An underground facility called HADES (High-Activity Disposal Experimental Site) excavated at 223-m depth close to the city of Mol was constructed in order to study the feasibility of High Level radioactive Waste (HLW) disposal in the Boom clay layer. Since construction in 1980, many experimental investigations have been made under both laboratory and field conditions, showing on the whole satisfactory capacities of sealing and healing under complex thermal, hydraulic, mechanical (THM) loading.

The thermo-mechanical behaviour of Boom clay has been widely investigated in the laboratory, mostly by performing temperature controlled high pressure triaxial tests (Baldi et al. 1988, 1991, Neerdael et al. 1992, De Bruyn and Thimus 1995, Bernier et al. 1995, Del Olmo et al., 1996; Belanteur et al. 1997, Sultan et al., 2002). Dilation was observed when heating the soil at low isotropic stresses or high over-consolidation ratio (OCR). Under high isotropic stresses (or low OCRs), heating first induced dilation and then contraction. Physically, saturated clays expand when heated. As the thermal expansion coefficient of water is much larger than that of the surrounding soil skeleton, the resulting restriction on water expansion yields an increase in pore water pressure (Britto et al. 1989). As a result consolidation takes place, leading to thermal contraction. The thermal consolidation of Boom clay was investigated by Delage et al., (2000). The effects of temperature on shear strength properties have been found to be strongly dependent on the volume changes induced by heating (Cui et al. 2000, Sultan et al. 2000). It has been shown that the thermal softening related to the decrease of water-clay interactions due to heating can be globally balanced by the volume decrease due to thermal consolidation. As far as constitutive modelling is concerned, the first elasto-plastic model taking into account the effects of temperature was reported by Hueckel and Baldi (1990) (see also Hueckel and Borsetto 1990). The model was afterwards modified by Picard (1994) who adopted a simpler exponential expression for the variations of isotropic yield stress with temperature. Based on a deeper examination of the experimental results from various authors (Paaswell 1967, Campanella and Mitchell 1968, Plum and Esrig 1969, Demars and Charles 1982, Houston and Lin 1987, Baldi et al. 1988, 1991, Eriksson 1989, Tidfors and Sällfors 1989, Towhata et al. 1993), Cui et al. (2000) proposed a multi-mechanism model able to account better for the effect of OCR on the volume change behaviour of clays under heating.

Time-dependent phenomena are significant in deep clays in many underground geotechnical problems (Giraud & Rousset, 1996; Bastiaens *et al.*, 2007) in relation to both the effects of pore water dissipation and of the soil skeleton viscosity. Towhata *et al.* (1993), Akagi & Komiya (1995) and Shimizu (2003) observed that heating accelerated both the consolidation and the secondary compression of saturated clays. Based on experimental data, Burghignoli *et al.* (2000) suggested that a better understanding of the thermomechanical behaviour of clays could be achieved by explicitly accounting for the viscosity of the soil skeleton.

Few experimental results concerning the effect of temperature on the time-dependent behaviour of deep clays are available in the literature. De Bruyn & Thimus (1996) performed an undrained heating test on Boom clay at a constant heating rate of 0.5°C/min from 30°C to 75 °C and then kept the temperature constant at 75°C. They recorded a continuous increase in pore-water pressure, even under constant temperature conditions. Unfortunately, the high heating rate applied (0.5 °C/min) and the short time duration after heating (about 45 min) did not permit clear identification of the creep phenomenon.

The present work aims at investigating the time-dependent behaviour of Boom clay under both thermal and mechanical loading. A high pressure thermal triaxial device was developed for this purpose (allowing confining and deviator stresses up to 32 MPa, and temperature control between 25 °C and 100 °C). Samples were saturated under the in-situ mean effective stress conditions to avoid any swelling. The tests started with constant rate thermal and/or mechanical consolidation and ended with isobar heating and/or isothermal compression, at a constant rate or in steps. Based on the experimental results, the effects of loading rate, heating rate and temperature on the volume change behaviour are analysed.

## 2. Material and sample preparation

The Boom clay specimens were extracted at a depth of 223 m in the Underground Research Facility (URF) of Mol, Belgium. As noted by Horseman et al. (1987) and commented on by Burland (1990), the Boom clay is geologically lightly overconsolidated. However, the yield stress of Boom clay is larger than the pre-consolidation pressure owing to creep and diagenesis, resulting in a significant overconsolidation ratio OCR = 2.4. Table 1 (Decleer *et al.*, 1983; Al-Mukhtar *et al.*, 1996) presents the mineralogical composition of Boom clay. The clay fraction is dominant (50% to 62% < 2 μm). The geotechnical properties are shown in Table 2. Note that the natural water content measured on samples varies in a wide range between 25% and 30% although *in-situ* measurements in the URF gave an average value of between 24.5% and 25.5%. It should be mentioned that the natural water contents measured in the laboratory are strongly dependent on the quality of the samples sealing and of the age of the samples. The samples tested by Coll (2005) had a water content of between 22% and 24% whereas the samples studied in this work ranged from 19.5% to 21.6% (see Table 4). The plasticity index of Boom clay is quite high, being between 37% and 50%, making Boom clay a high plasticity clay according to Casagrande's classification. Table 2 also gives values of the angle of internal shearing resistance as well as the hydraulic conductivity confirming that Boom clay is a stiff clay of low hydraulic conductivity ($10^{-12}$ m/s).

Soil blocks taken during the excavation of the connecting gallery were cut into 85 × 40 × 40 mm sub-blocks which were then machined using a lathe to a cylinder of 38-mm diameter and 76-mm height. A synthetic water having a chemical content similar to the in-situ pore water (see Table 3) was used in the tests in order to minimize any chemical perturbation.

## 3. Experimental methods

### *3.1. Testing device*
A testing device allowing high-pressure triaxial compression tests at controlled temperature was developed (Figure 1). The system comprises an auto-compensated triaxial cell (Sultan 1997) able to sustain a confining pressure of 60 MPa. The soil specimen is installed between two initially dry porous stones to avoid any swelling. The specimen is enveloped by two membranes (0.6-mm thick and 37 mm in diameter) in order to avoid any leakage due to membrane tearing. Both the cell base and the cap are connected to two ducts that allow proper saturation of the top and bottom porous stones by flushing. A pressure transducer was installed outside the cell for the pore water pressure measurement at the top.

The confining pressure and the back pressure are applied by two pressure/volume controllers (GDS). A heating coil is placed on the outer wall of the cell and the temperature is measured by a thermocouple installed inside the cell, near the soil sample. A temperature regulator is used for the temperature control. The back pressure is measured by the GDS connected to the sample and by the pressure transducer connected to the top of it. During a test, the soil volume changes are measured by monitoring the water volume changes of the back-pressure GDS. The ambient temperature is controlled at 25 °C. All water ducts are made of steel and are immersed in water baths regulated at 25 °C.

### *3.2. Experimental procedure*

As explained above, careful saturation of the samples was carried out prior to testing, as described in Figure 2. After mounting the soil sample on the pedestal with dry porous stones, a confining pressure of 100 kPa was applied at a temperature regulated at 25 °C. Once the cell volume change (measured by the GDS) stabilised, the confining pressure was increased at a low rate of 0.5 kPa/min to 2500 kPa with valves opened and no pore water pressure changes recorded by the bottom GDS and the top pressure transducer. In such conditions, the sample is subjected to a mean effective stress close to that in-situ, estimated at 2.5 MPa by Horseman et al. (1993) with a $K_0 = 0.8$ (see also Delage et al. 2007).

Once the soil volume stabilised under 2500 kPa, the sample was put in contact with water under a pressure of 100 kPa at the bottom after flushing the porous stone with the top drainage valve still opened (no pressure recorded by the top pressure transducer). During this stage, the volume of water penetrating the sample was monitored by the back-pressure GDS while sample volume change was monitored by the confining pressure GDS. After about 8 days, water injection was performed at the top of the sample in a similar fashion (giving rise to a 100 kPa pressure recorded by the pressure transducer as indicated in Figure 2). In this stage, the confining pressure was also increased by 100 kPa. Afterwards, simultaneous increases of both the confining pressure and the back pressure were then applied in steps of 250 kPa (except for the first step from 100 kPa to 250 kPa) till reaching a confining pressure of 3500 kPa and a back pressure of 1000 kPa. The difference between the confining pressure and the back pressure was hence kept constant and equal to 2500 kPa, the mean in-situ effective stress. The Skempton coefficient $B$ was determined during the last step with average values close to 0.92 obtained.

### *3.3. Testing programme*

Once saturated, the samples were subjected to thermal and/or mechanical loading by increasing temperature or confining stress either at a constant rate or in steps.
Table 4 and Figure 3 summarise the heating and loading paths followed. Table 4 also gives the void ratio values at the initial state and after saturation process described above (water injection and loading to 2.5 MPa mean effective stress). In tests 3, 8, 11, and 13, the samples were loaded at 25 °C from an initial effective mean stress of 2.5 MPa to respectively 10, 3.5, 4, and 3.5 MPa. In test 13, the sample was then heated from 25 to 76 °C at a mean stress (confining pressure minus back pressure) of 3.5 MPa, whereas in test 11, the sample was heated to 67 °C under 4.0 MPa. In tests 7, 12, and 15, the samples were initially heated at 76 °C (80 °C in case of test 12) under 2.5 MPa. The sample in test 12 was then cooled from 80 to 70 °C and reloaded to 3.0 MPa. The sample in test 15 was loaded from 2.5 to 3.5 MPa at 76 °C.

In the isothermal tests at 25°C, volume changes were satisfactorily measured from the water exchanges monitored by the GDS for back pressure. This was no longer possible during the heating test because of the thermal dilation of the soil constituents as well as that of the system (water in the tubings and in the cell). The correction proposed by Campanella & Mitchell (1968) was applied. This correction considers the free water in the soil and gives the following expression of the volumetric strain $\varepsilon_v$ obtained during a drained heating test:

$$\varepsilon_v = [\Delta V_{dr} - (\alpha_w V_w + \alpha_s V_s)\Delta T]/V \qquad (1)$$

where $\Delta V_{dr}$, $V_w$, $V_s$ and $V$ are respectively the volume of drained water, pore water, solid skeleton and soil sample; $\Delta T$ is the temperature change; $\alpha_w$ and $\alpha_w$ are respectively the thermal expansion coefficients of water and of the solid skeleton. In low-porosity plastic clays, Baldi *et al.* (1988) suggested considering the effect of adsorbed water by using the double-layer diffuse theory. They considered that the pressure of adsorbed water is exponentially decreasing with the distance between the water molecule and the clay platelet. Delage *et al.* (2000) analysed the experimental data obtained on Boom clay and concluded that the simpler method of Campanella & Mitchell (1968) was satisfactory for describing the volume change of Boom clay in a heating test. This method was adopted in this work for calculating volume changes.

During sample saturation, soil volume changes were monitored by the confining-pressure GDS, particularly under constant confining pressure.

## 4. Experimental results

### *4.1. Soil volume change during the saturation stage*

Figure 4 presents the swelling observed in tests 3, 7, 8, 13 when samples submitted to a 2500 kPa confining pressure were put in contact with water. Swelling measurements were deduced from the water exchanges between the cell and the confining pressure GDS. Note that the time duration in each test depended upon the swelling evolution. This stage was stopped when swelling started stabilising. Stabilisation seems to be reached in tests 8 and 13 but not in the two other tests. The magnitude of swelling observed ranges from 0.8% (test 8) to 2.3% (test 13). This is suspected to be related to the difference in sample disturbance during preparation. Examination of the curves of Figure 4 shows a comparable swelling rate for all samples.

The observed swelling is comparable to that obtained by Coll (2005) who saturated natural Boom clay samples (initial water content $w = 22 - 24\%$) under an effective isotropic stress of 2.3 MPa and observed a magnitude of swelling smaller than 4%. Interestingly, Coll (2005) also found after a period of time of two weeks a continuous swelling at a rate of $9\times10^{-9}$ s$^{-1}$ very close to that of test 13 ($8\times10^{-9}$ s$^{-1}$).

It appeared to be difficult to saturate the sample by water injection from the base because, even after one week, no water flux was observed on the top of the sample. Zhang et al. (2007)

observed the same phenomenon on Opalinus clay and attributed it to the pore clogging due to soil swelling.

Figure 5 presents the changes in water volume monitored by the confining-pressure GDS ($dV_{CP}$) and by the back-pressure GDS ($dV_{BP}$), together with the changes of the confining pressure ($P_{CP}$) in test 3. The negative sign for $dV$ means that the controllers lost some water (volume decrease for $dV_{CP}$ and water adsorption for $dV_{BP}$). During swelling, water entered the sample ($dV_{BP}$ increased). It was observed that the volume of water adsorbed by the soil was much larger than the soil's swelling, the latter being close to zero. This difference shows that the adsorbed water mainly served to fill the existing voids occupied by air, thus saturating the soil.

At each subsequent increase of back pressure and confining pressure, $dV_{CP}$ shows an instantaneous decrease followed by an apparent stabilization. As expected, the recorded instantaneous $dV_{CP}$ decrease corresponded exactly to the deformability of the system evaluated during a calibration test using a metallic dummy sample. The water penetration from the back-pressure GDS ($dV_{BP}$) seems to be more progressive: during the first stage at $P_{CP}$ = 2600 kPa, about 140 hours were needed to reach stabilisation. The stabilisation time is shorter with higher pressure. During the two last steps, the stabilisation was almost instantaneous, indicating that the soil was fully saturated with all voids filled by water.

### *4.2. Mechanical and thermal volumetric behaviour*

Figure 6 presents the results of test 3. After being saturated, the soil sample was loaded and unloaded at 25 °C following a cycle: 2.5 – 10.0 – 2.5 MPa. As proposed by Sultan (1997), a loading and unloading rate of 0.5 kPa/min was applied. The maximum mean stress (confining pressure minus back pressure, $p'$) of 10.0 MPa was maintained for 100 hours. After unloading to 2.5 MPa, the pressure was kept constant for 150 hours. The volumetric strain shows that the soil sample kept compressing at the end of the loading path (10.0 MPa) and kept swelling at the end of the unloading path (2.5 MPa), evidencing a significant time-dependent behaviour. This ongoing variation of volumetric strain can also be observed in Figure 7 that presents the volumetric strain and void ratio versus either $p'$ (Figure 7a) or $\log p'$ (Figure 7b). Examination of the volumetric strain coefficient $\frac{d\varepsilon_v}{d\log t}$ (Figure 6) by considering the results obtained 24 hours after the end of loading or unloading showed that $\frac{d\varepsilon_v}{d\log t} = 8.45\times10^{-2}$ at $p'$ = 10.0 MPa and $\frac{d\varepsilon_v}{d\log t} = -12.49\times10^{-2}$ at $p'$ = 2.5 MPa. Note that the decision of considering the results obtained after a period of time of 24 hours has been made so as to define a criterion applicable to all tests and to allow for comparison between the results.

Figure 8 presents the results, in graphs of $e$-log$t$, of all the tests following either a heating or a loading path (tests are summarised in Table 4 and Figure 3). In each test, all the results are presented under an isobar and isothermal condition following a compression or a heating phase. By only considering the results in the interval 24 hours – 100 hours, the value of $\frac{d\varepsilon_v}{d\log t}$ can be estimated for certain curves with mechanical or thermal loading in slope or in steps. The obtained values are presented in Table 5, together with the corresponding test conditions.

The isothermal consolidation can be investigated by considering the loading stages in tests 3, 8, 11, 12, 13 and 15. Figure 8*c* shows two relatively similar *e*-log*t* curves obtained at 25 °C on the one hand after a step loading compression (2.5 to 3.0 MPa) and on the other hand after a compression at constant rate of stress (3.0 to 3.5 MPa, 0.2 kPa/min, test 8). The values of $\frac{d\varepsilon_v}{d\log t}$ look quite small on the two curves, close to zero. On the contrary, a much larger volumetric strain rate coefficient was obtained in test 3 ($\frac{d\varepsilon_v}{d\log t}$ = 8.45×10$^{-2}$) after loading to 10.0 MPa at a constant rate of stress of 0.5 kPa/min (Figure 8a). This appears to be due to several factors: i) a higher loading rate, ii) a higher applied stress (10.0 MPa against 3.5 MPa), iii) a larger loading interval (2.5 MPa - 10 MPa against 3 – 3.5 MPa) because of the likely time dependency of pore water pressure generation. Furthermore, as opposed to the stress of 3.5 MPa, the stress of 10.0 MPa is higher than the pre-consolidation pressure estimated at 5 MPa (see Coll, 2005 and Delage *et al.*, 2007). This can be also a reason for the larger volumetric strain rate coefficient observed in test 3.

A lower stress rate was applied in test 11 (0.1 kPa/min) when increasing *p'* from 3.5 to 4.0 MPa at 25 °C (Figure 8*d*). At constant *p'* = 4 MPa, the soil volume continued to decrease with a volumetric strain rate coefficient of $\frac{d\varepsilon_v}{d\log t}$ = 0.43×10$^{-2}$, a value quite small similar to that observed in test 8. The same observation is obtained from test 13 (Figure 8*f*): $\frac{d\varepsilon_v}{d\log t}$ = 0.56×10$^{-2}$ after compression at a stress rate of 0.1 kPa/min from 2.5 to 3.0 MPa.

Examination of compression tests at high temperatures shows that, in test 12 (70 °C, Figure 8*e*), a $\frac{d\varepsilon_v}{d\log t}$ value of 0.85×10$^{-2}$ was obtained after compression from 2.5 to 3.0 MPa. In test 15 (76 °C, Figure 8*g*), a slightly larger value of $\frac{d\varepsilon_v}{d\log t}$ = 0.93×10$^{-2}$ was obtained at 3.0 MPa. This increase can be on one hand related to the effect of temperature and on the other hand to the effect of void ratio (0.515 in test 12 against 0.435 in test 15). Under a 3.5 MPa pressure, the value of $\frac{d\varepsilon_v}{d\log t}$ rises to 3.77×10$^{-2}$ (Figure 8*g*), showing a significant stress effect.

Figure 9 shows the variation of void ratio versus time, in *e*-log*t* graph, after isothermal compressions at 25 °C and 70-80 °C. It can be observed that the curves at high temperatures are steeper, showing that the volumetric strain rate coefficient is larger as compared with those obtained at 25 °C. This observation is confirmed in Figure 10 where all the values of $\frac{d\varepsilon_v}{d\log t}$ are presented as a function of temperature. The $\frac{d\varepsilon_v}{d\log t}$ value of test 3, although performed at 25 °C, is significantly higher than the others. As mentioned above, this can be explained by the relatively high stress rate applied during compression (0.5 kPa/min), the high stress level (10 MPa) and the large loading interval.

The thermal consolidation behaviour can be analysed by considering the heating stages in test 7 (Figure 8*b*), test 11 (Figure 8*d*), test 12 (Figure 8*e*), test 13 (Figure 8*f*) and test 15 (Figure 8*g*). In most cases the curves exhibit a compression behaviour under heating (tests 7, 11, and 13). However an expansion was observed in tests 12 and 15 under low temperature steps: 40 and 55 °C in the case of test 12 and 39.4 °C in the case of test 15. Interestingly, this expansion was observed at a low pressure of 2.5 MPa.

Figure 8 shows that the volumetric compression rate is dependent on the temperature value, the higher the temperature, the higher the compression rate. Taking test 13 (Figure 8*f*) as an example, the curve becomes steeper when temperature is higher. An exception is observed in test 11 (Figure 8*d*) in which the curve at 39.5 °C appears to be steeper than that at 52.6 °C. This can be explained by the fact that heating from 39.5 to 52.6 °C was performed at a constant rate of 0.28 °C/h, whereas heating from 25 to 39.5 °C was applied in one step (see Table 5). In the former, part of the consolidation would have been achieved during constant heating whereas in the latter, the $e$-log$t$ curve reflects the whole consolidation process.

Figure 11 presents the variation of $\frac{d\varepsilon_v}{d\log t}$ after the heating phases in test 7 (constant rate heating at 0.28 °C/h) and tests 11, 13, 15 (step heating), as a function of temperature. A clear increase in $\frac{d\varepsilon_v}{d\log t}$ with increasing temperature is observed. Furthermore, the $\frac{d\varepsilon_v}{d\log t}$-$T$ relationship can be described by an exponential function. Note that this observation is made for quite a low stress range; experimental data from a higher range are clearly needed to confirm this point.

Figure 12 presents the variation of $\frac{d\varepsilon_v}{d\log t}$ versus temperature for all the tests performed. The significant scatter observed is related to the influence of stress level, void ratio, thermal or mechanical loading rate, loading mode and loading interval, In spite of this, a clear trend showing an increase in $\frac{d\varepsilon_v}{d\log t}$ with increasing temperature can be observed.

## 5. Discussion

### *5.1. Sample saturation*

As proposed by Delage *et al.* (2007), the soil specimen was submitted to a confining pressure close to the in-situ effective mean stress (2.5 MPa) prior to any contact with water. This procedure significantly decreased the swelling observed when the soil was put in contact with water. A similar observation was also made by Coll (2005) during sample saturation under an effective isotropic stress of 2.3 MPa. This specific saturation procedure reduces the soil swelling and thus any related microstructure changes. Nevertheless, as seen in Figure 4, swelling was not completely eliminated and it can be as large as 2.3%. In addition, the magnitude of swelling was not the same for all the samples tested. One could conclude in a first approach that the applied 2.5 MPa effective stress was not high enough. However, after comparing the initial water contents of the tested samples (19.5 to 21.6%) with that of soil in situ (24 to 25%, see Delage *et al.,* 2007), it seems probable that soil drying during sample transport, storage, trimming in lathe machine etc., would be the cause of the swelling observed. Note also that the stress applied in the cell was isotropic, which is not the case in-situ.

*5.2. Mechanical volumetric behaviour*

The ongoing contraction under constant pressure observed once loading is completed in isothermal compression tests can be related either to a too high stress rate that did not allow total pore pressure dissipation or to creep and related microstructure reorganisation. After an isothermal (25°C) compression at a high stress rate up to a higher stress (0.5 kPa/min from 2.5 to 10.0 MPa), a much larger volumetric strain rate coefficient ($\frac{d\varepsilon_v}{d\log t}$ = 8.45×10$^{-2}$) than that obtained after a compression at lower rates (0.2 kPa/min and 0.1 kPa/min) to lower stress (3.0, 3.5 or 4 MPa) was observed ($\frac{d\varepsilon_v}{d\log t}$ close to zero). At first sight, this difference is believed to be due to higher loading rate, higher stress level and larger loading interval. The results also suggest that at a low temperature of 25°C, loading to a stress lower than the pre-consolidation pressure (5 MPa) at a rate lower than 0.2 kPa/min results in negligible volumetric strain rate coefficient. In other words, there was complete dissipation of pore water pressure in this case. This observation obviously requires confirmation by performing a test with pore water pressure measurements within the soil sample.

*5.3. Thermal volumetric behaviour*

Thermal expansion of the samples was observed under low effective stress (2.5 MPa) and low temperatures. By contrast, thermal consolidation was observed under higher effective pressures (3.5 and 4.0 MPa) and higher temperatures. Taking the pre-consolidation pressure of Boom clay equal to 5.0 MPa, an over-consolidation ratio (*OCR*) of 2 is obtained with p' = 2.5 MPa. Hence, the heating tests in which contraction and thermal consolidation was observed (3.5 MPa and 4.0 MPa) correspond to smaller values of *OCR* (1.40 and 1.25 respectively). This is consistent with the results of Plum & Esrig (1969) who observed a volume decrease on a remoulded illite in the normally consolidated state (*OCR* = 1) or slightly over-consolidated state (*OCR*<1.7). Conversely, they reported a volume increase with *OCR* higher than 1.7. A similar *OCR* effect was also identified on Boom clay by Baldi *et al.* (1988), Cui *et al.* (2000) and Sultan *et al.* (2002).

As Table 5 indicates, a heating rate of 0.28 °C/h was applied in test 7 (from 25 to 76 °C) and in test 11 (from 40 to 53 °C). In test 7 a $\frac{d\varepsilon_v}{d\log t}$ value of 7.48×10$^{-2}$ was recorded whereas in test 11 a much smaller value of 0.78×10$^{-2}$ was recorded. This difference clearly shows the effect of temperature on the volumetric strain rate coefficient. The relationship between $\frac{d\varepsilon_v}{d\log t}$ and temperature level can be described by an exponential function (Figure 11).

## 6. Conclusions

An experimental investigation was carried out on Boom clay using a testing system allowing high-pressure isotropic compression tests at controlled temperatures. Seven tests were performed by following various loading or heating modes (constant rate or in steps) and various loading and heating paths. Emphasis was put on the volumetric strain rate coefficient evolution at constant effective pressure and constant temperature. The following conclusions can be drawn from the results obtained.

i) Because soil samples have lost some water during transport, storage, preparation etc., water infiltration under a mean effective stress close to the in-situ value (2.5 MPa) can give rise to some swelling, with a maximum observed value of 2.3%. It is however expected that after this swelling the soil sample state approached that of the soil in situ because for clays as stiff as Boom clay, the volume change due to wetting-drying can be considered approximately recoverable.

ii) A significant effect of stress and temperature on the volumetric strain rate coefficient was observed. Regarding the effect of the loading rate, the results showed that at a low temperature of 25°C, loading to a stress lower than the pre-consolidation pressure at a rate lower than 0.2 kPa/min resulted in negligible volumetric strain rate coefficients for Boom clay. That suggests a complete dissipation of pore water pressure in this case. This point should obviously be checked by performing more tests under similar conditions (same void ratio, larger range of loading rates and heating rates) with pore water pressure measurements within the soil samples.

iii) When heated, Boom clay exhibited either a volume increase or a volume decrease, depending on the *OCR* value and temperature range. Thermal consolidation occurred under small *OCR* conditions whereas thermal expansion occurred under large *OCR* conditions.

iv) After low-rate loading or heating stages, the ongoing volume change under constant pressure or constant temperature, respectively, can be mostly related to creep. This is not the case in quick loading or heating tests, particularly for step loading or heating tests. In the latter case, the separation between consolidation and creep makes it necessary either to perform some back analysis of the results or to carry out some thermal or mechanical tests with pore pressure measurements in the soil sample.

## 7. Acknowledgements

ONDRAF/NIRAS (The Belgian Agency for Radioactive Waste and Enriched Fissile Materials) is gratefully acknowledged for its financial support. This work is part of the PhD thesis prepared at Ecole des Ponts – ParisTech by the second author. The financial support of Ecole des Ponts – ParisTech is also acknowledged.

## 8. References

Akagi, H. & Komiya, K. (1995). Constant rate of strain consolidation properties of clayey soil at high temperature. *Proc. Int. Symp. Compression and Consolidation of Clayey Soils – IS –Hiroshima '95*. (H. Yoshikuni and O. Kusakabe (eds)). pp. 3 – 8.

Al-Mukhtar, M., Belanteur, N., Tessier, D., & Vanapalli, S.K. (1996). The fabric of a clay soil under controlled mechanical and hydraulic stress states. *Applied Clay Science* **11**, Nos. 2-4, 99-115.

Baldi, G., Hueckel, T. & Pellegrini, R. (1988). Thermal volume changes of the mineral-water system in low-porosity clay soils. *Can. Geotech. J.* **25**, No. 4, 807 – 825.

Baldi, G., Hueckel, T., Peano, A. & Pellegrini, R. (1991). *Developments in modelling of thermo-hydro-geomechanical behaviour of boom clay and clay-based buffer materials (vol. 2)*. Commission of the European Communities, Nuclear Science and Technology EUR 13365/2.

Bastiaens, W., Bernier, F. & Li, X.L. (2007). An Overview of Long-Term HM Measurements around HADES URF. *Proc. Intern. Symp. Multiphysics coupling and long term behaviour in rock mechanics*, Liège, 2006, May 9-12, 15 – 26.

Belanteur, N., Tacherifet, S. & Pakzad, M. (1997). Étude des comportements mécanique, themo-mécanique et hydro-mécanique des argiles gonflantes et non gonflantes fortement compactées. *Revue Française de Géotechnique* **78**, 31 – 50.

Bernier, F., Volckaert, G., Alonso, E. & Villar, M. (1995). Suction controlled experiments on Boom clay. *International Workshop on Hydro-Thermo-Mechanics of Engineered Clay Barriers and Geological barriers*.

Britto A.M., Savvidou C., Maddocks D.V., Gunn M.J. & Booker J.R. (1989). Numerical and centrifuge modelling of coupled heat flow and consolidation around hot cylinders buried in clay. *Géotechnique* **39**, No. 1, 13–25.

Burghignoli, A., Desideri, A. & Miliziano, S. (2000). A laboratory study on the thermomechanical behaviour of clayey soils. *Can. Geotech. J.* **37**, No. 4, 764 – 780.

Burland, J. (1990). On the compressibility and shear strength of natural clays. *Géotechnique* **40**, No. 3, 329-378.

Campanella, R. G. & Mitchell, J. K. (1968). Influence of temperature variations on soil behavior. *Journal of the Soil Mechanics and Foundations Division*, *ASCE*, **94**, No. 3, 709-734.

Coll, C. (2005). *Endommagement des roches argileuses et perméabilité induite au voisinage d'ouvrages souterrains*. Ph.D. thesis, Université Joseph Fourier-Grenoble 1, France.

Cui, Y.J, Sultan, N. & Delage, P. (2000). *Propriétés thermomécaniques des milieux granulaires. Chapitre VI-2 in "Mécanique des milieux granulaires".* Edition HERMES (ISBN 2-7462-0289-1), 323-363.

Cui, Y. J., Sultan, N. & Delage, P. (2000). A thermomechanical model for clays. *Can. Geotech. J.* **37**, No. 3, 607 – 620.

De Bruyn, D. & Thimus, J.F. (1995). The influence of anisotropy on clay strength at high temperature. *Proceedings of the 11th European Conference on Soil Mechanics and Foundation Engineering*, Copenhagen, Vol. 3, pp. 37-42.

De Bruyn, D. & Thimus, J.F. (1996). The influence of temperature on mechanical characteristics of Boom clay: the results of an initial laboratory programme. *Engineering Geology* **41**, Nos. 1 – 4, 117 – 126.

Decleer, J., Viane, W. & Vandenberghe, N. (1983). Relationships between chemical, physical and mineralogical characteristics of the Rupelian Boom clay, Belgium. *Clay Minerals* **18**, 1 – 10.

Dehandschutter, B., Vandycke, S., Sintubin, M., Vandenberghe, N. & Wouters, L. (2005) Brittle fractures and ductile shear bands in argillaceous sediments: inferences from Oligocen Boom clay (Belgium). *Journal of Structural Geology* **27**, No. 6, 1095 – 1112.

Del Olmo, C., Fioravante, V., Gera, F., Hueckel, T., Mayor, J. C. & Pellegrini, R. (1996). Thermomechanical properties of deep argillaceous formations. *Engineering Geology* **41**, Nos. 1 – 4, 87 – 101.

Delage, P., Sultan, N. & Cui, Y.J. (2000). On the thermal consolidation of Boom clay. *Can. Geotech. J.* **37**, No. 2, 343 – 354.

Delage, P., Le, T.T., Tang A.M., Cui, Y.J. & Li, X.L. (2007). Suction effects in deep Boom clay samples. *Géotechnique* **57**, No. 2, 239 – 244.

Demars, K.R. & Charles, R.D. (1982). Soil volume changes induced by temperature cycling. Can. Geotech. J. **19**, 188-194.

Eriksson, L.G. (1989). Temperature effects on consolidation properties of sulphide clays. *Proceedings of the 12th International Conference on Soil Mechanics and Foundation Engineering*, 3, 2087-2090.

Giraud, A. & Rousset, G. (1996). Time-dependent behavior of deep clays. *Engineering Geology* **41**, Nos. 1-4, 181 – 195.

Horseman, S. T., Winter, M. G. & Entwistle, D.C. (1987). *Geotechnical characterisation of Boom clay in relation to disposal of radioactive waste.* Luxembourg: Office for Official Publications of the European Communities.

Horseman, S. T., Winter, M. G. & Entwistle, D. C. (1993). Triaxial experiments on Boom clay. *The Engineering Geology of Weak Rock*, Balkema, Rotterdam, 36-43.

Houston, S. L. & Lin, H.D. (1987). Thermal consolidation model for pelagic clays. *Marine Geotechnology* **7**, 79-98.

Hueckel T. & Baldi, G. (1990). Thermoplasticity of saturated clays: experimental constitutive study. *Journal of Geotechnical Engineering* **116**, No. 12, 1778-1796.

Hueckel, T. & Borsetto, M. (1990). Thermoplasticity of saturated soils and shales: Constitutive equations. *Journal of Geotechnical Engineering, ASCE* **116**, No. 12, 1765-1777.

Neerdael, B., Beaufays, R., Buyens, M., De Bruyn D. & Voet M. (1992). *Geomechanical behaviour of Boom clay under ambient and elevated temperature conditions.* Commission of the European Communities, Nuclear Science and Technology EUR 14154, 108 pages.

Paaswell, R.E. (1967). Temperature effects on clay soil consolidation. *Journal of Soil Mechanics and Foundation Engineering, ASCE.* **93**, SM3, 9-22.

Picard, J. (1994). *Ecrouissage thermique des argiles saturées : application au stockage des déchets radioactifs*. Ph.D. thesis, Ecole Nationale des Ponts et Chaussées, Paris, 283 pages.

Plum, R.L. & Esrig, M.I. (1969). Some temperature effects on soil compressibility and pore water pressure. *Effects of Temperature and heat on Engineering Behavior of Soils, Special Report 103, Highway Research Board*, 231 – 242.

Shimizu, M. (2003). Quantitative assessment of thermal acceleration of time effects in one-dimensional compression of clays. *Proc. 3$^{rd}$ Int. Symp. Deformation Characteristics of Geomaterials*, Lyon, 479 – 487.

Sultan, N. (1997). Etude du comportement thermo-mécanique de l'argile de Boom: expériences et modélisation. Ph.D. thesis, Ecole Nationale des Ponts et Chaussées, Paris.

Sultan, N., Delage, P & Cui, Y.J. (2000). Comportement thermomécanique de l'argile de Boom. *Comptes-Rendus de l'Académie des Sciences, Série II B-Mécanique* **328**, No. 6, 457-463.

Sultan, N., Delage, P. & Cui, Y. J. (2002). Temperature effects on the volume change behaviour of Boom clay. *Engineering Geology* **64**, Nos. 2 – 3, 135 – 145.

Tidfors, M. & Sällfors, G. (1989). Temperature effect on preconsolidation pressure. *Geotechnical Testing Journal* **12,** No. 1, 93-97.

Towhata, I., Kuntiwattanakul, P., Seko, I. & Ohishi, K. (1993). Volume change of clays induced by heating as observed in consolidation tests. *Soils and Foundations* **33**, No. 4, 170-183.

Zhang, C.-L., Rothfuchs, T., Su, K. & Hoteit, N. (2007). Experimental study of the thermo-hydro-mechanical behaviour of indurated clays. *Physics and Chemistry of the Earth* **32**, Nos. 8-14, 957 – 965.

|  | Decleer *et al.* (1983) (%) | Al-Mukhtar *et al.* (1996) (%) |
|---|---|---|
| ▪ Clay minerals | 50 | 62 |
|    Illite | 12 | 16 |
|    Kaolinite | 5 | 13 |
|    Smectite | 33 | 33 |
| ▪ Quartz | 35 | 20-25 |
| ▪ Calcite, Dolomite | 1 | - |
| ▪ Pyrite | 1 | 4-5 |
| ▪ Feldspar |  |  |
|    Microcline | 9 | 4-5 |
|    Plagioclase | 4 | 4-5 |

**Table 1. Mineralogical composition**

|  | Belanteur *et al.* (1997) | Dehandschutter *et al.* (2005) |
|---|---|---|
| Specific gravity | 2.67 |  |
| Bulk density (Mg/m$^3$) |  | 1.9 |
| Liquid limit $w_L$ (%) | 59-76 | 70 |
| Plastic limit $w_P$ (%) | 22-26 | 25 |
| Plastic index $I_P$ | 37-50 | 45 |
| Water content (%) |  | 25-30 |
| Natural porosity (%) |  | 35 |
| Poission's ratio |  | 0.4 |
| Angle of internal shearing resistance (°) |  | 18 |
| Hydraulic conductivity (m/s) |  | $10^{-12}$ |

**Table 2. Geotechnical properties**

| Salt | Concentration (mg/l) |
|---|---|
| $CaCO_3$ | 5000.0 |
| $NaHCO_3$ | 1 170.0 |
| $H_3BO_3$ | 43.0 |
| KCl | 25.0 |
| $MgCl_2 \cdot 6H_2O$ | 22.0 |
| NaF | 11.0 |
| NaCl | 10.0 |
| $Na_2SO_4$ | 0.3 |

**Table 3. Salts composition in pore-water**

| Test No. | $e_i$ | $w_i$ (%) | $e_c$ | Heating and loading paths followed |
|---|---|---|---|---|
| 3 | 0.62 | 21.6 | 0.59 | p' = 2.5 – 10.0 - 2.5 MPa (0.5 kPa/min) at 25 °C |
| 7 | 0.59 | 19.6 | 0.55 | p'= 2.5 MPa; T° = 25 – 76 °C at (0.28 °C/h) |
| 8 | 0.63 | 19.6 | 0.61 | p' = 2.5 – 3.0 MPa at 25 °C<br>p' = 3.0 – 3.5 MPa (0.2 kPa/min) at 25 °C |
| 11 | 0.61 | 19.8 | 0.57 | p' = 2.5 – 4.0 MPa (0.1 kPa/min) at 25 °C<br>T° = 25 - 40 - 53 - 67 °C at p' = 4.0 MPa |
| 12 | 0.63 | 20.0 | 0.65 | p' = 2.5 MPa. T°C = 25 - 40 - 55 - 70 - 80 - 70 °C<br>p' = 2.5 – 3.0 MPa at 70 °C<br>p' = 2.5 – 3.0 MPa |
| 13 | 0.59 | 19.5 | 0.58 | p' = 3.0 – 3.5 MPa (0.1 kPa/min) at 25 °C<br>T° = 25 - 39.5 - 48.7 - 57.9 - 67 - 76.1 °C at p' =3.5 MPa<br>p' = 2.5 MPa. T° = 25 - 39.4 - 48.8 - 58 - 67 - 76.3 °C |
| 15 | 0.60 | 21.6 | 0.57 | p'= 2.5 – 3.0 – 3.5 MPa at 76.3 °C |

Note: $e_i$: initial void ratio; $e_c$: void ratio after the saturation and the loading to 2.5 MPa; $w_i$: initial water content

**Table 4. Summary of experiments carried out**

| Test No | Loading and heating paths followed | p' (MPa) | T (°C) | $d\varepsilon_v/d(\log t) \times 100^{(a)}$ |
|---|---|---|---|---|
| Test 3 | Loading 0.5 kPa/min from 2.5 to 10.0 MPa | 10.0 | 25 | 8.45 |
| Test 3 | Unloading 0.5 kPa/min from 10.0 to 2.5 MPa | 2.5 | 25 | 12.41 |

| Test 7 | Heating 0.28 °C/h from 25 to 76 °C | 2.5 | 76 | 7.48 |
| --- | --- | --- | --- | --- |
| Test 8 | Loading from 2.5 to 3.0 MPa | 3.0 | 25 | 0.21 |
| Test 8 | Loading 0.2 kPa/min from p'=3.0 to 3.5 MPa | 3.5 | 25 | 0.46 |
| Test 11 | Loading 0.1 kPa/min from p'=2.5 to 4.0 MPa | 4.0 | 25 | 0.43 |
| Test 11 | Heating from 25 to 40 °C | 4.0 | 40 | 0.19 |
| Test 11 | Heating 0.28 °C/h from 40 to 53 °C | 4.0 | 53 | 0.78 |
| Test 11 | Heating from 53 to 67 °C | 4.0 | 67 | 5.53 |
| Test 12 | Heating from 55 to 70 °C | 2.5 | 70 | 3.87 |
| Test 12 | Heating from 70 to 80 °C | 2.5 | 80 | 28.90 |
| Test 12 | Cooling from 80 to 70 °C | 2.5 | 70 | 28.19 |
| Test 12 | Loading from 2.5 to 3.0 MPa | 3.0 | 70 | 0.85 |
| Test 13 | Loading from 2.5 to 3.0 MPa | 3.0 | 25 | 0.09 |
| Test 13 | Loading 0.1 kPa/min from 3.0 to 3.5 MPa | 3.5 | 25 | 0.56 |
| Test 13 | Heating from 25 to 40 °C | 3.5 | 39.5 | 0.31 |
| Test 13 | Heating from 40 to 49 °C | 3.5 | 48.7 | 1.02 |
| Test 13 | Heating from 49 to 58 °C | 3.5 | 57.9 | 3.42 |
| Test 13 | Heating from 58 to 67 °C | 3.5 | 67 | 7.17 |
| Test 13 | Heating from 67 to 76 °C | 3.5 | 76.1 | 13.51 |
| Test 15 | Heating from 40 to 50 °C | 2.5 | 48.8 | 0.61 |
| Test 15 | Heating from 50 to 58 °C | 2.5 | 58 | 1.44 |
| Test 15 | Heating from 58 to 67 °C | 2.5 | 67 | 2.91 |
| Test 15 | Heating from 67 to 76.3 °C | 2.5 | 76.3 | 7.57 |
| Test 15 | Loading from 2.5 to 3.0 MPa | 3.0 | 76.3 | 0.93 |
| Test 15 | Loading from 3.0 to 3.5 MPa | 3.5 | 76.3 | 3.77 |

[a] Note: determined by considering the results in the interval between 24 hours and 100 hours, after the end of loading, unloading or heating.

**Table 5. Summary of testing results: volumetric strain rate coefficient at an isobar and isothermal state calculated 24 hours following a heating or loading path**

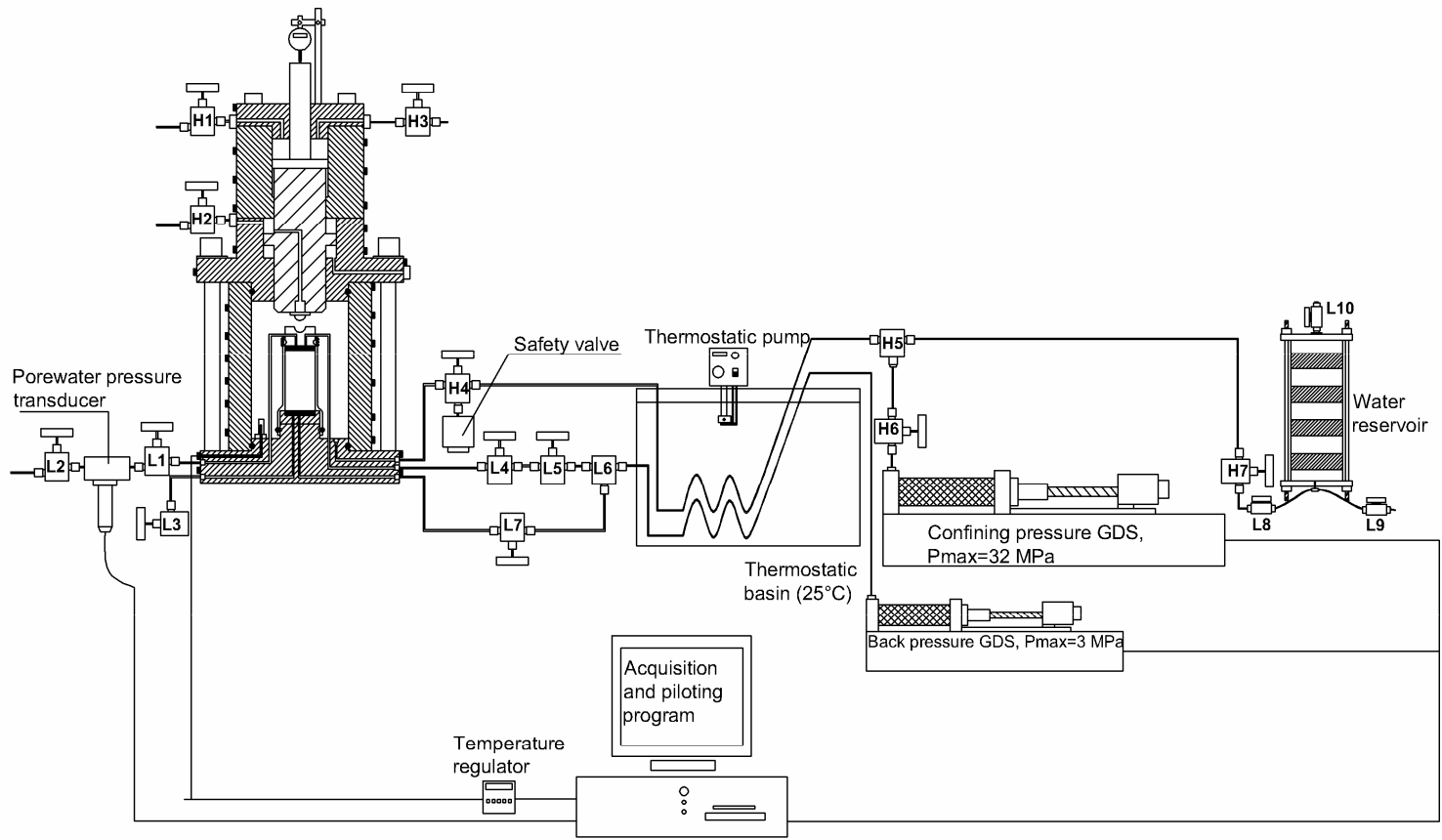

Note: valves represented by "L" are low-pressure valves whereas valves represented by "H" are high-pressure valves.

**Figure 1. Experimental system for high pressure and temperature tests.**

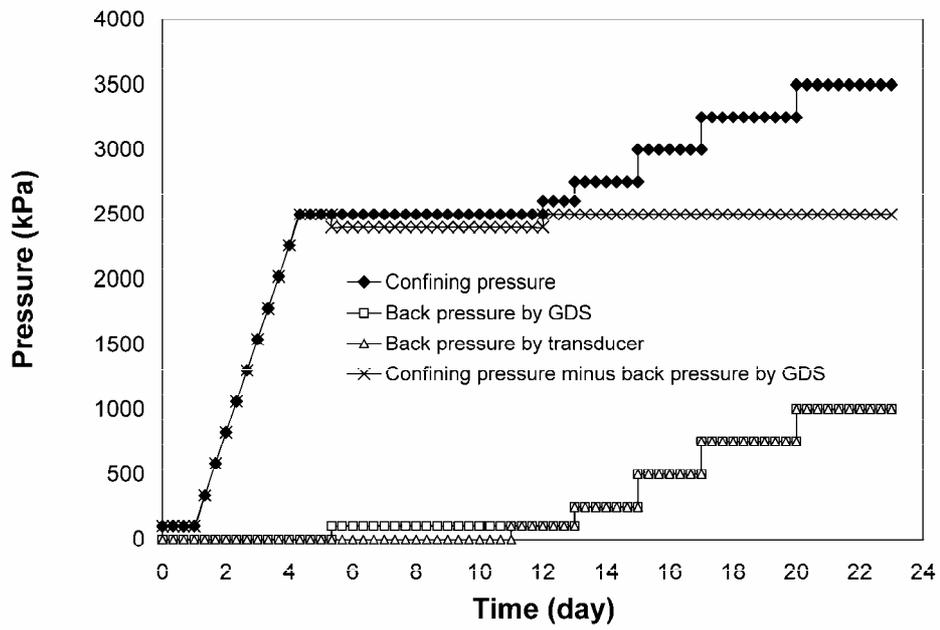

**Figure 2. Stress path followed during the saturation phase.**

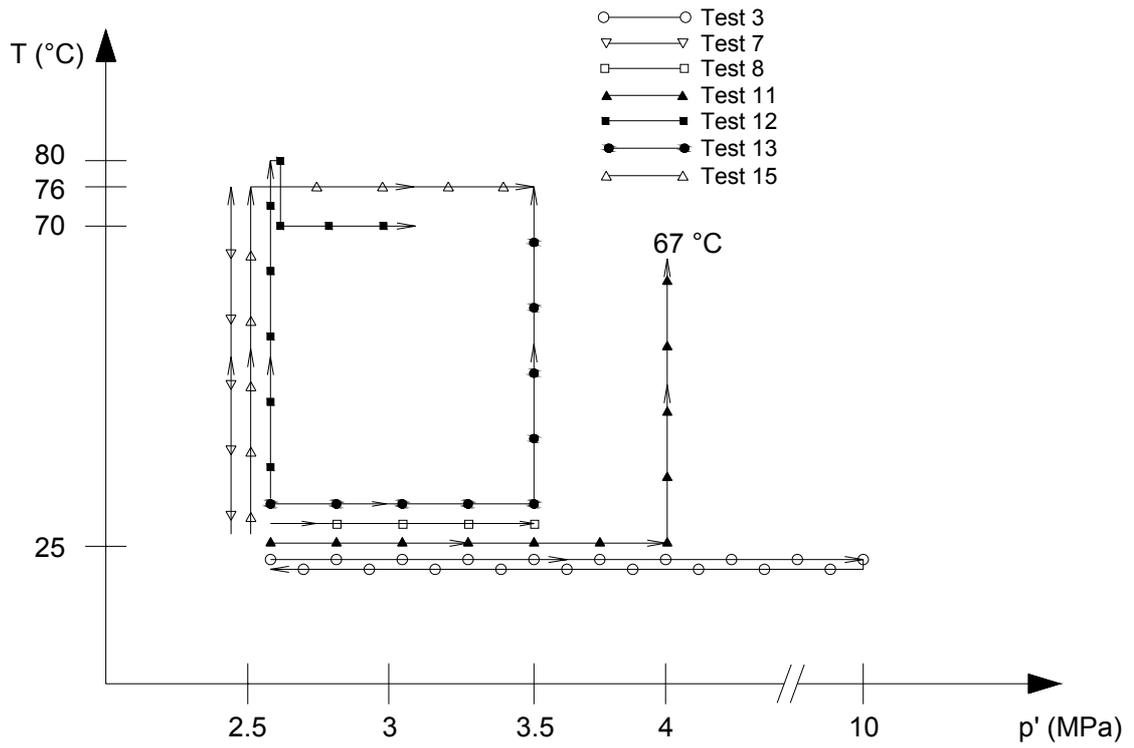

Figure 3. Loading and heating paths followed.

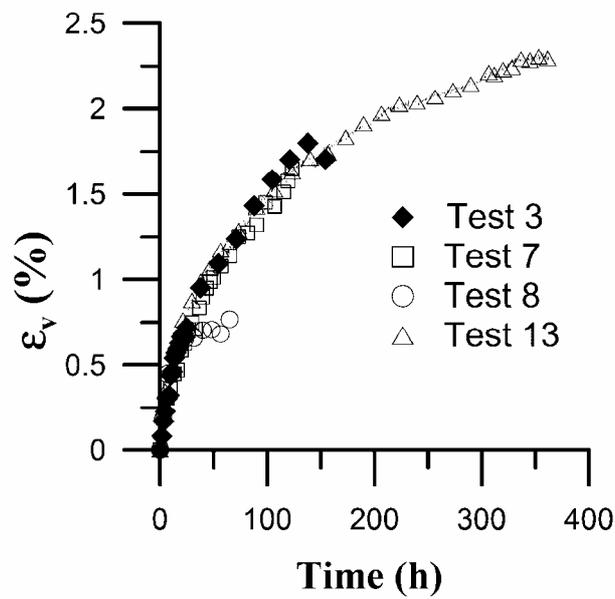

Figure 4. Swelling following the sample-water contact.

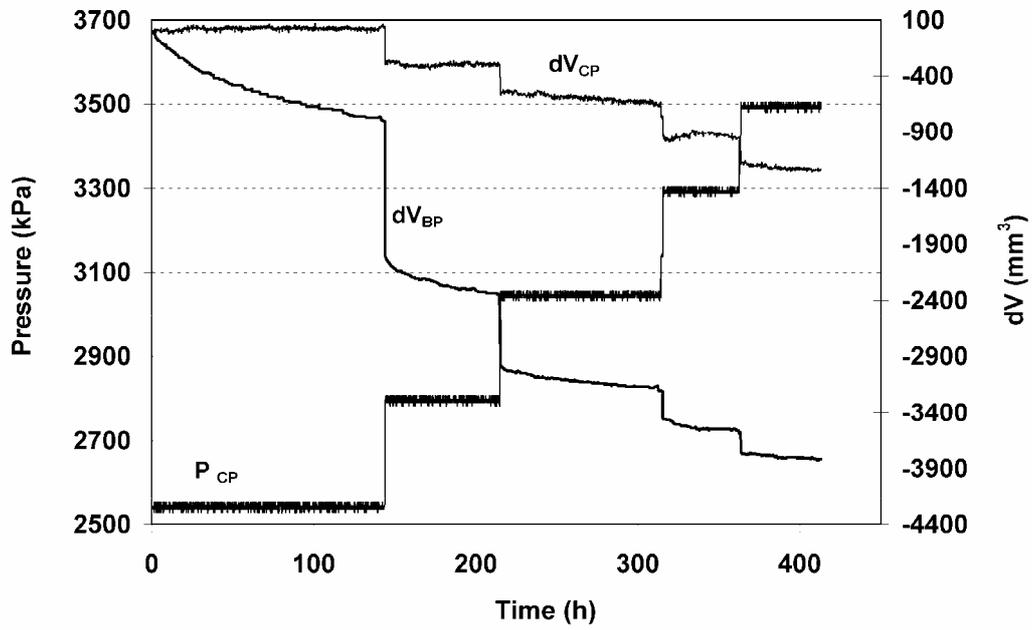

Figure 5. Volumetric changes monitored by the GDS applying the confining pressure (dV$_{CP}$) and the GDS applying the back pressure (dV$_{BP}$) (Effective mean stress kept equal to 2.5 MPa, test 3).

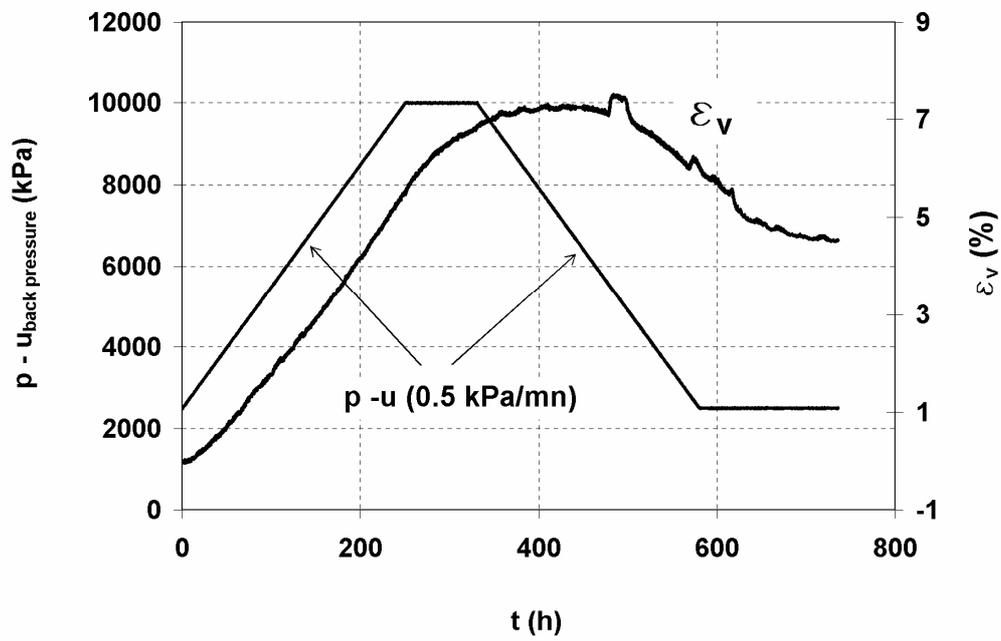

**Figure 6. Volumetric strain and effective mean stress versus time during a loading and unloading test (test 3)**

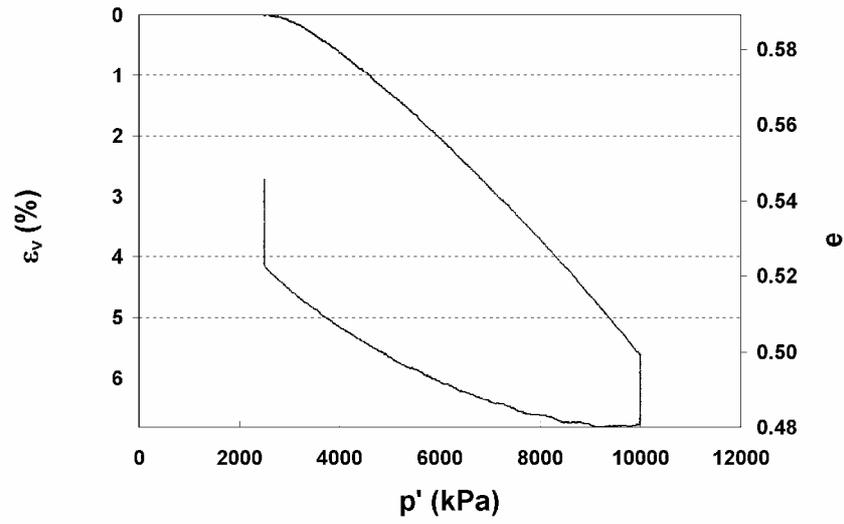

(a)

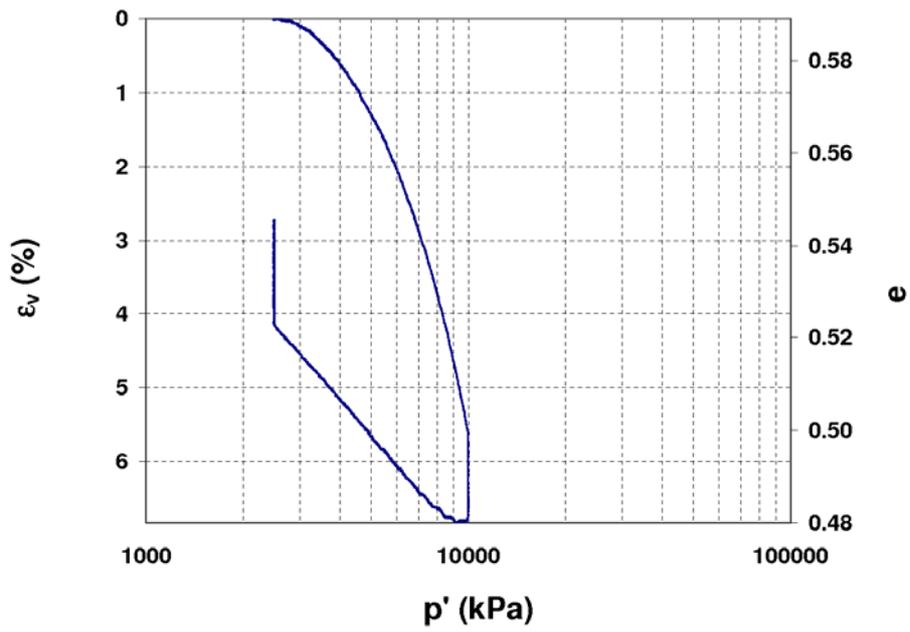

(b)

**Figure 7. Volumetric strain during a loading and unloading test (test 3).**

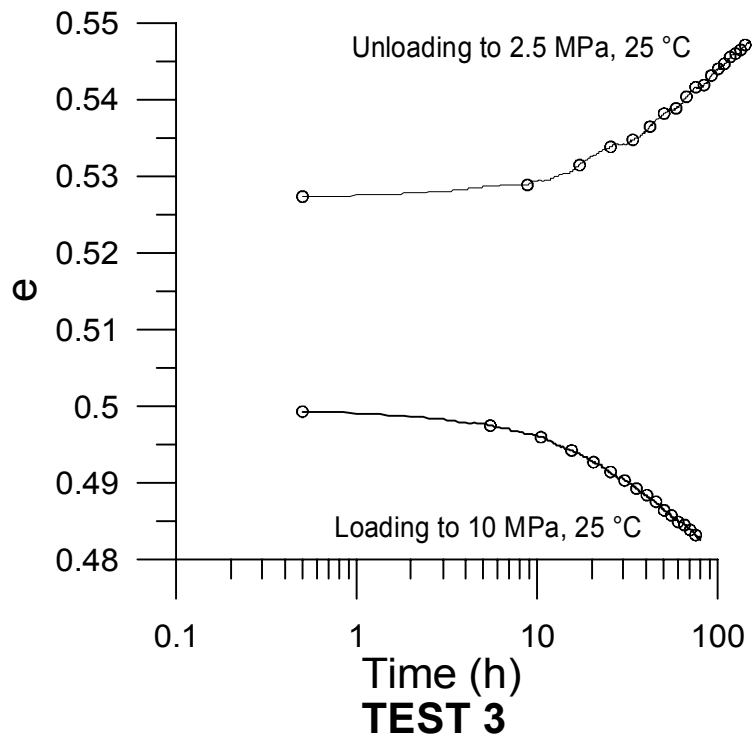

**(a)**

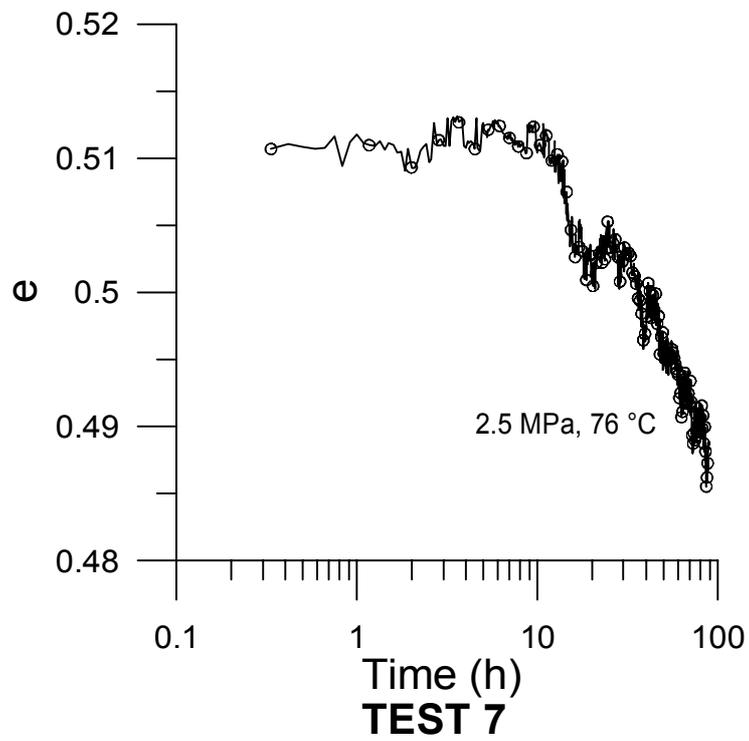

**(b)**

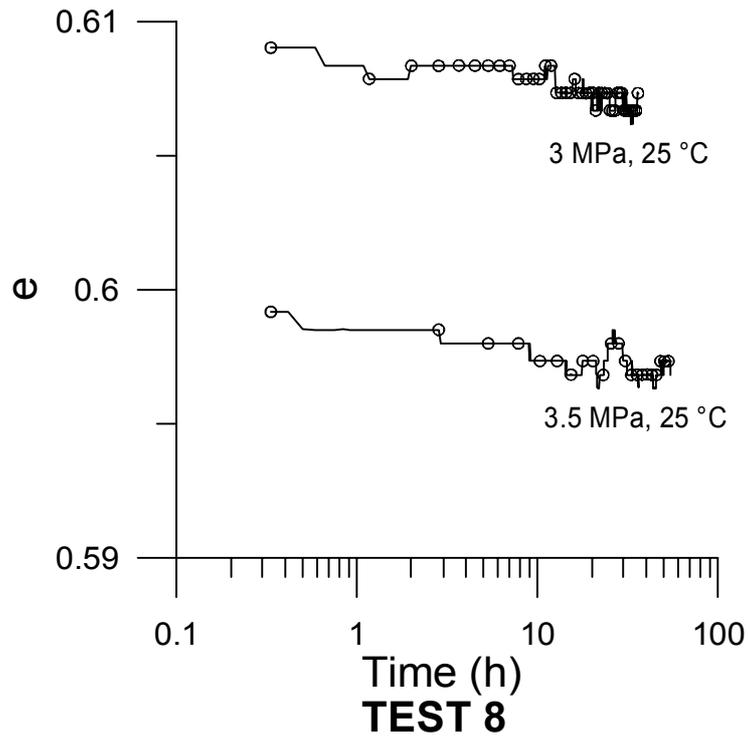

**(c)**

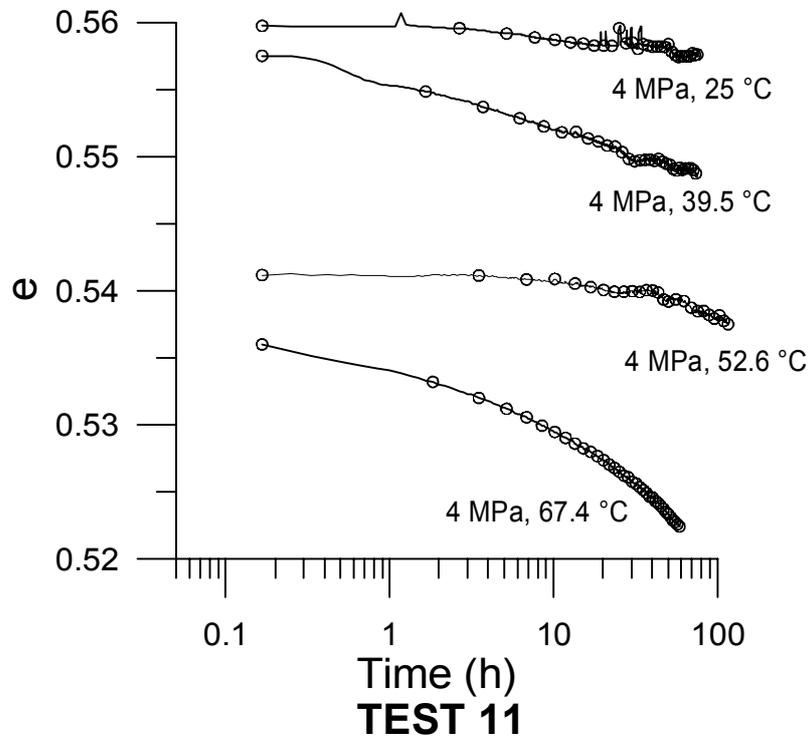

**(d)**

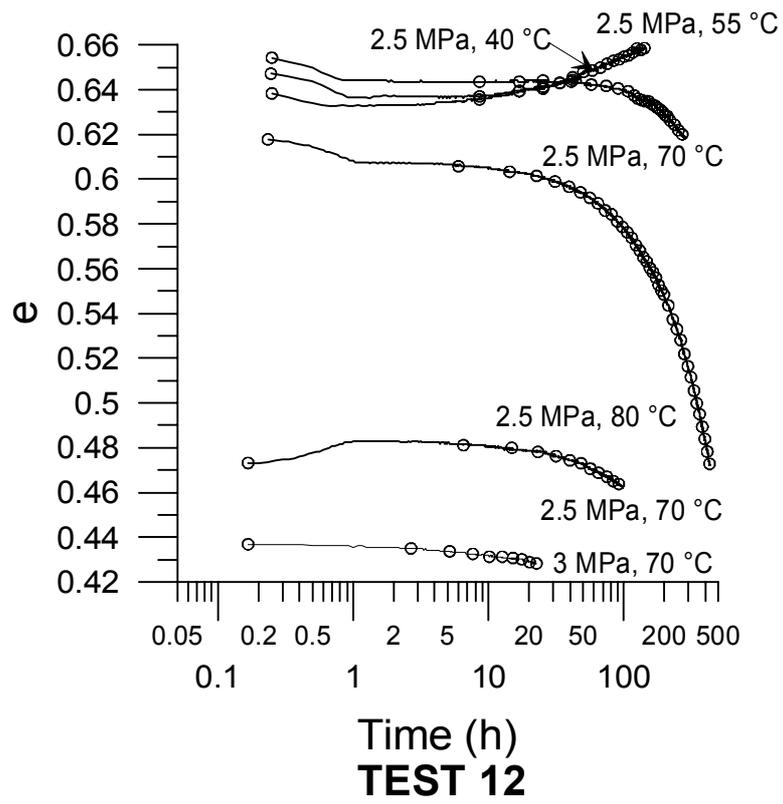

**(e)**

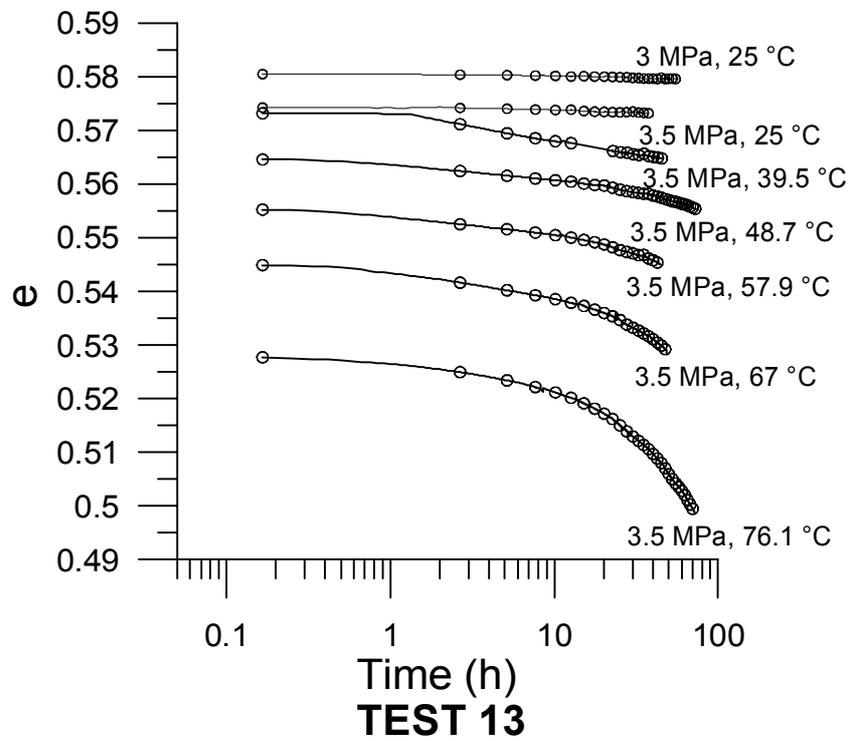

**(f)**

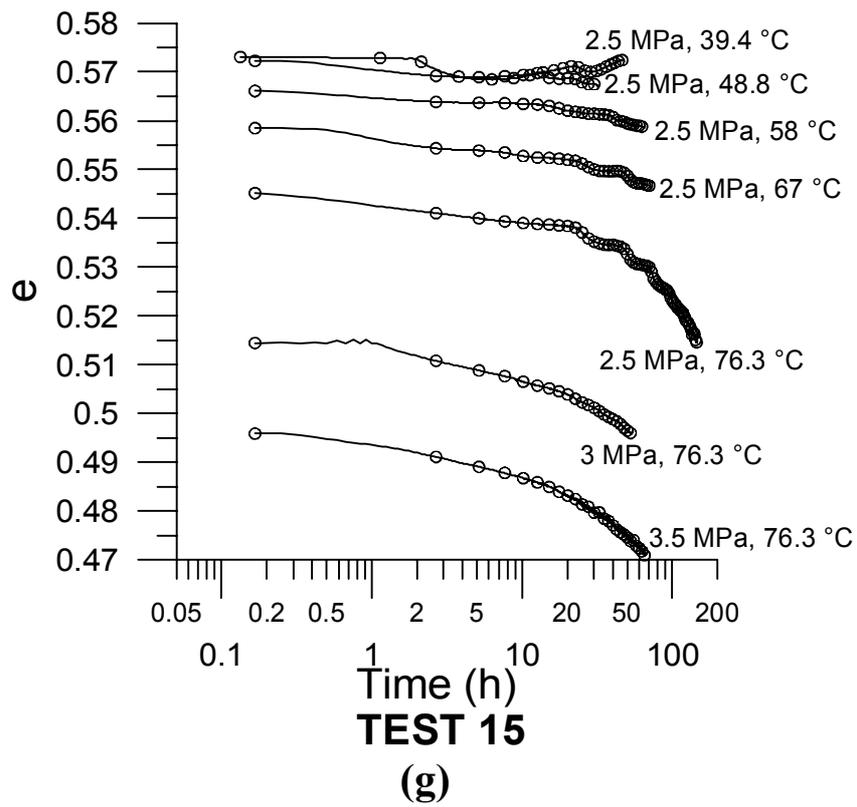

**TEST 15**

**(g)**

**Figure 8. Testing results: void ratio variation at constant temperature and constant effective pressure following a heating or loading path**

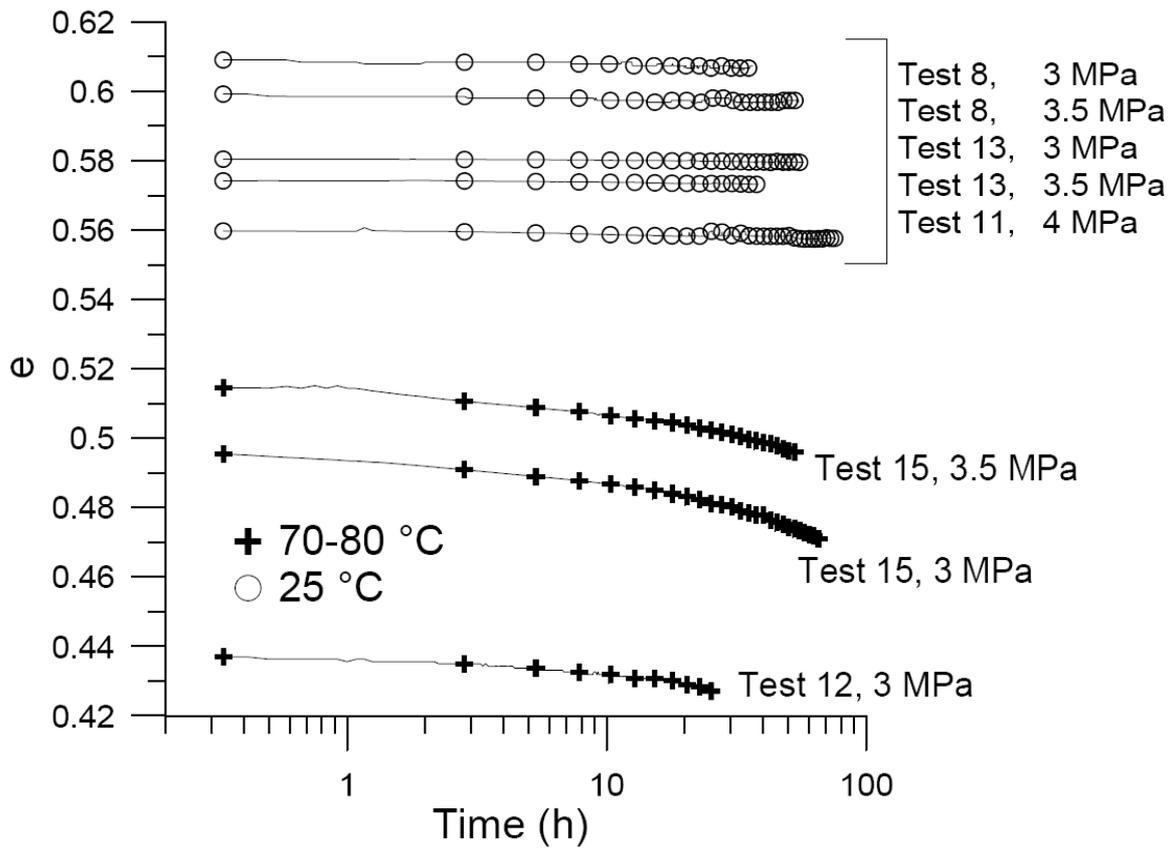

Figure 9. Void ratio variation at the end of compression test (constant effective mean stress and temperature).

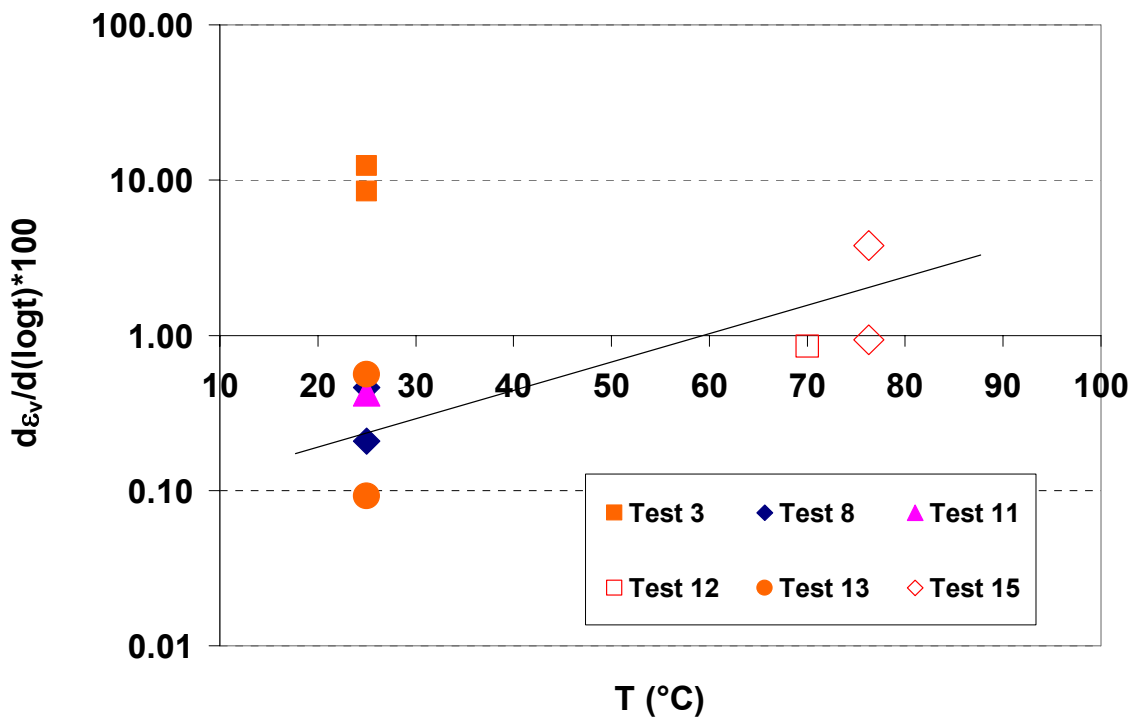

Figure 10. Consolidation rate versus temperature obtained following compression tests.

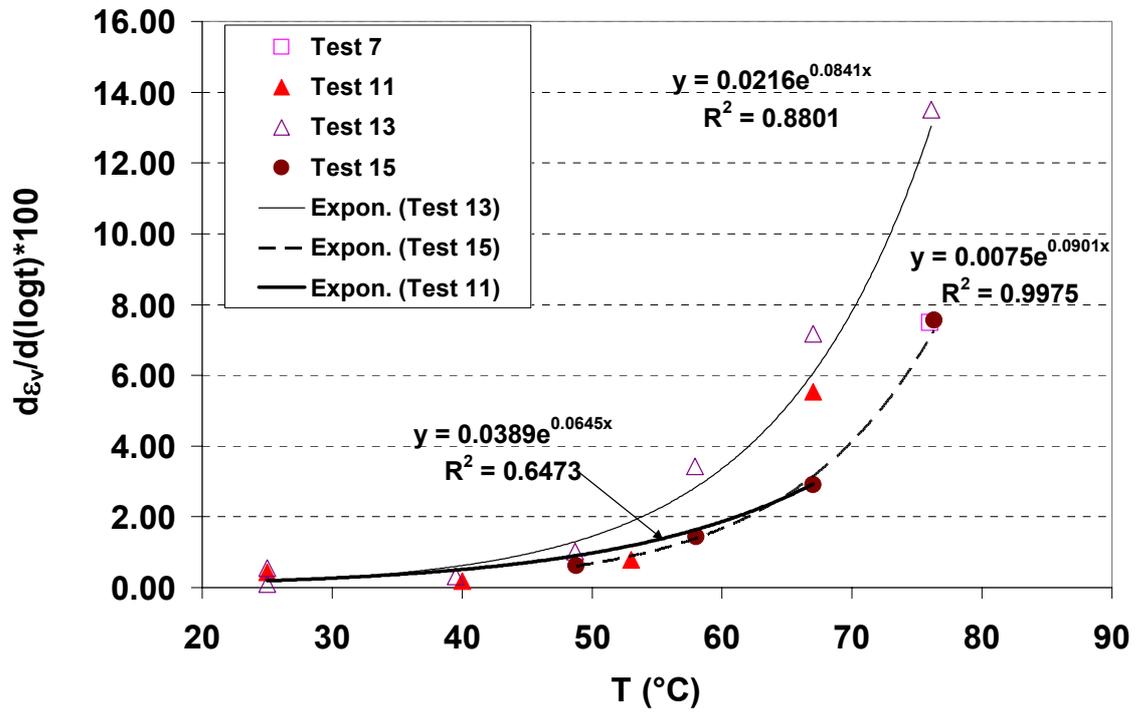

**Figure 11. Consolidation rate versus temperature of different heating tests.**

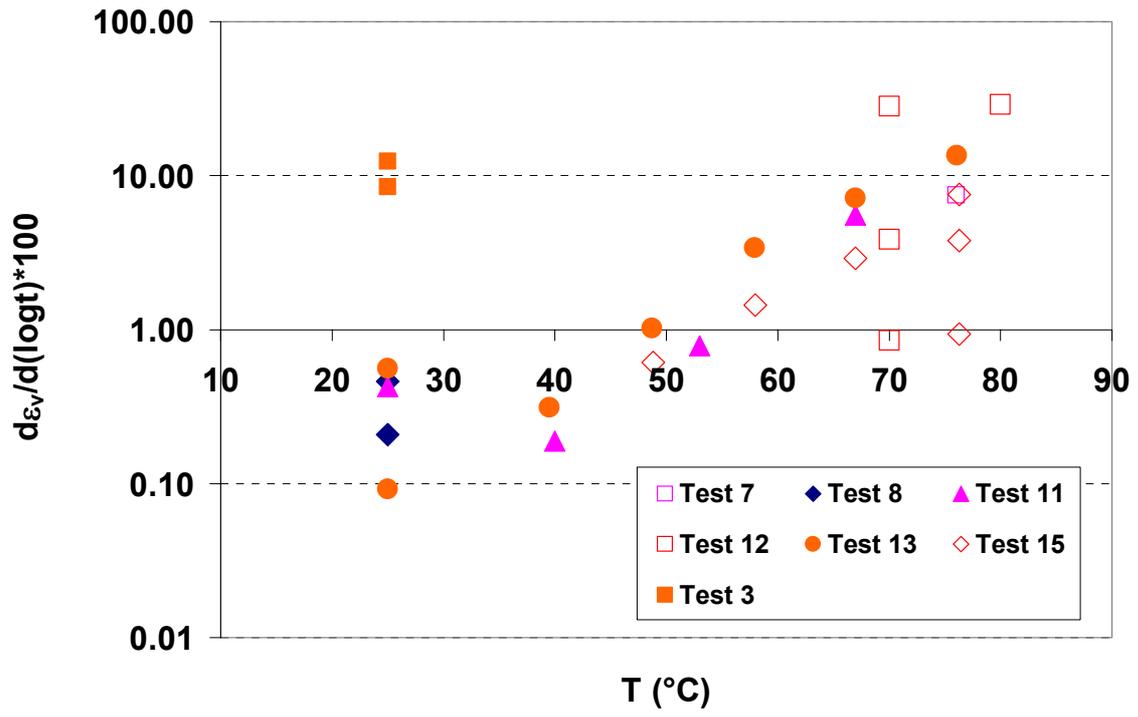

**Figure 12. Consolidation rate versus temperature of all the tests.**